\documentclass[12pt]{iopart}
\usepackage{epsfig}

\begin{document}
\jl{1}
\bibliographystyle{plain}

\title[Modulated magnetic structures]{Spatially modulated magnetic structures in thin films}
\author{W Selke\S,~ D Catrein\S,~and ~M Pleimling\dag\ddag}
\address{\S\ Institut f\"ur Theoretische Physik, Technische
Hochschule, D--52056 Aachen, Germany}
\address{\dag\ Institut f\"ur Theoretische Physik 1, Universit\"at Erlangen-N\"urnberg, D--91058 Erlangen, Germany}
\address{\ddag\  Laboratoire de Physique des Mat\'eriaux\footnote[5]{ Laboratoire associ\'e au CNRS UMR 7556},
Universit\'e Henri Poincar\'e Nancy I, B.P. 239, 
F--54506 Vand{\oe}uvre l\`es Nancy Cedex, France}

\begin{abstract}
The axial next--nearest neighbour Ising (ANNNI) model is 
studied for thin films of up to $L= 10$ layers, with
a distinct phase diagram for each film thickness. The
systematics of the ordered phases, as obtained from
mean--field theory, Monte Carlo simulations, and low
temperature expansions, is discussed. Results are compared to those for
the ANNNI model in the limit $L \longrightarrow \infty$.
\end{abstract}
\maketitle

In recent years magnetism in thin films of a few atomic
layers has attracted much interest \cite{Poul}. However, studies on
the influence of the layer thickness, $L$, on spatially modulated
magnetic structures seem to be very scarce \cite{Hafner}.

In the following, we
shall deal with this aspect by analysing phase diagrams
of the axial next--nearest neighbour Ising (ANNNI)
model \cite{Selke,Yeo,Pleim} 
on a simple cubic lattice (setting the
lattice constant equal to one) consisting of rather few, but
large layers. The competing interactions are supposed to
be ferromagnetic between
neighbouring spins in each layer, $J_0>0$, as well as between neighbouring
spins in adjacent layers, $J_1>0$, and to be
antiferromagnetic, $J_2<0$, between
axial next--nearest neighbour spins, distinguishing one of the three
cubic axes, say, the $z$--axis. The resulting magnetic structures
may exhibit non--trivial spatial modulations along the $z$--axis. In the
case of indefinitely many and indefinitely large layers, the
model displays a phase diagram with a plenitude of commensurate
phases, including those springing from
the multiphase point at zero temperature and
those emerging from structure combination
branching processes at finite temperatures, as well as incommensurate
structures, see the reviews \cite{Selke,Yeo,Pleim}.    
 
Here we shall consider thin films of
$L$ layers, with {\it free} boundary conditions
at the surface layers. Note that the
ANNNI model on finite lattices, connecting the
first and last layers by {\it periodic} boundary conditions, has
been studied quite extensively before, attempting to predict properties
for $L \longrightarrow \infty$. The 
scope of the present study is, however, to determine
the phase diagram of the ANNNI model with {\it free} boundary conditions
at fixed film thickness. Attention may
be also drawn to recent work on the semi--infinite ANNNI model, analysing
critical surface properties near the Lifshitz point \cite{Binder}, which
partly motivated our investigation.
  
To establish the phase diagrams for
thin films, in particular for $L=3, 4, ...,10$, we used
mean--field theory, Monte
Carlo simulations and low temperature expansions. For
simplicity, we set $J_0= J_1$, both in the surface layers and
in the bulk of the film. In mean--field
approximation \cite{Bak,Dux,Jan}, we solved the set of
coupled non--linear
equations for the layer magnetizations, $m(z=1,...,L)$, by 
standard iteration
procedures starting from all possible combinations of fully
ordered, $m= 1$ or $m= -1$, or completely disordered, $m= 0$, layers, i.\ e.\
from, in principle,  $3^L$ distinct configurations. To identify the
stable phase, at given temperature, $\tau= k_B T/J_0$,
and ratio of competing interactions, $\kappa = - J_2/J_1$, we calculated the
free energy for all resulting structures. This approach
seems to be suitable to map correctly, in mean--field
approximation, the entire phase diagram. The
Monte Carlo simulations, using the single--spin flip
algorithm, were performed for lattices with $K \times K$ spins
in each layer connected by periodic boundary conditions (for thicker
films, an efficient layer flip algorithm may be useful \cite {Matsu}). To
circumvent finite size effects due to the
layer size, $K$ has to be chosen
sufficiently large. Actually, we varied $K$
from 10 to 100. To identify the phases springing directly from the 
degenerate ground states at $\kappa= 1/2$, low temperature
expansions to leading, first
order \cite{Fisher1,Fisher2} were found to suffice in most cases.    

At zero temperature, both for $L$ odd and even, the ferromagnetic state
is stable at $\kappa < 1/2$. For $L$ even, at $\kappa > 1/2$, the ground
state is the $\langle 2^{L/2} \rangle$ structure, where
two layers of "up" ("down") spins are followed by two layers of "down"
("up") spins, with totally ($L/2$) pairs of such layers (in
general, $\langle ...k l... \rangle$ refers to a phase at low temperatures
where a "$k$--band", i.\ e.\ spins
in $k$ adjacent layers being oriented (predominantly) in one
direction, say, "+", is followed by a $l$--band of
spins being oriented (predominantly) in the
opposite direction , say, "--"\cite{Fisher2}). For $L$ odd, structures
consisting of 2--bands and one 3--band have the lowest energy
at $1/2 < \kappa < 1$, while at $\kappa > 1$ the ground state
contains 2--bands with a 1--band at one of the surfaces (of
course, $\kappa= 1$ plays no special
role for the ANNNI model with 
$L \longrightarrow \infty$). The
transition line, at $\kappa \approx 1$, between
the corresponding phases is, at low temperatures, of first order.

The degeneracy, $D_L$, of the multiphase
point, at ($\kappa= 1/2$, $\tau= 0$), is related to the
Fibonacci sequence 
by $D_L= D_{L-2} + D_{L-1}$, with $D_2= D_3$= 2. We studied
the vicinity of that point by using low temperature
expansions \cite{Fisher1,Fisher2}. In the
limit $L \longrightarrow \infty$ \cite{Fisher1}, indefinitely
many periodic, commensurate phases spring directly from that
point, namely the ferromagnetic phase and phases in each of which 3--bands
are separated by a fixed number, $j$, of
2--bands, $j= 0, 1, 2, ... \infty$. In thin films, obviously only
a few phases may arise from the multiphase point, in
particular, the
ferromagnetic phase as well as additional phases formed mostly
by 2--bands and 3--bands. For $L$ even, the $\langle 2^{L/2} \rangle$ phase
will be among the additional phases, while for $L$ odd, all of these phases
will have at least one 3--band (when it is just one 3--band, then
that band will be preferably in the center of the film). Note that structures
consisting of 3--bands and a supplementary single 4--band or
a single 5--band at one of the
surfaces of the film occur when $L= 3n +1$ or
$L= 3n +2$, $n= 2, 3, ..$. Examples of those phases bordering, at
low temperatures, the ferromagnetic phase are 
the $\langle 43 \rangle$, $\langle 53 \rangle$, and
$\langle 433 \rangle$ phases. Again in contrast
to the case $L \longrightarrow \infty$, phases with 2--bands and
{\it pairs} of 3--bands may be stable as well; examples are the
$\langle 2332 \rangle$ and $\langle 332 \rangle$ phases.-- To
understand the phase diagram in the vicinity of
the multiphase point one should be aware of the fact that the stable
phases correspond to the
ground state structures with a low energy and a large
number of lowest--energy
excitations. The latter property favours 3--bands, where the spins
in both edge layers can be excited easily when both the preceeding and
the following band comprise at least two layers.

\begin{figure}
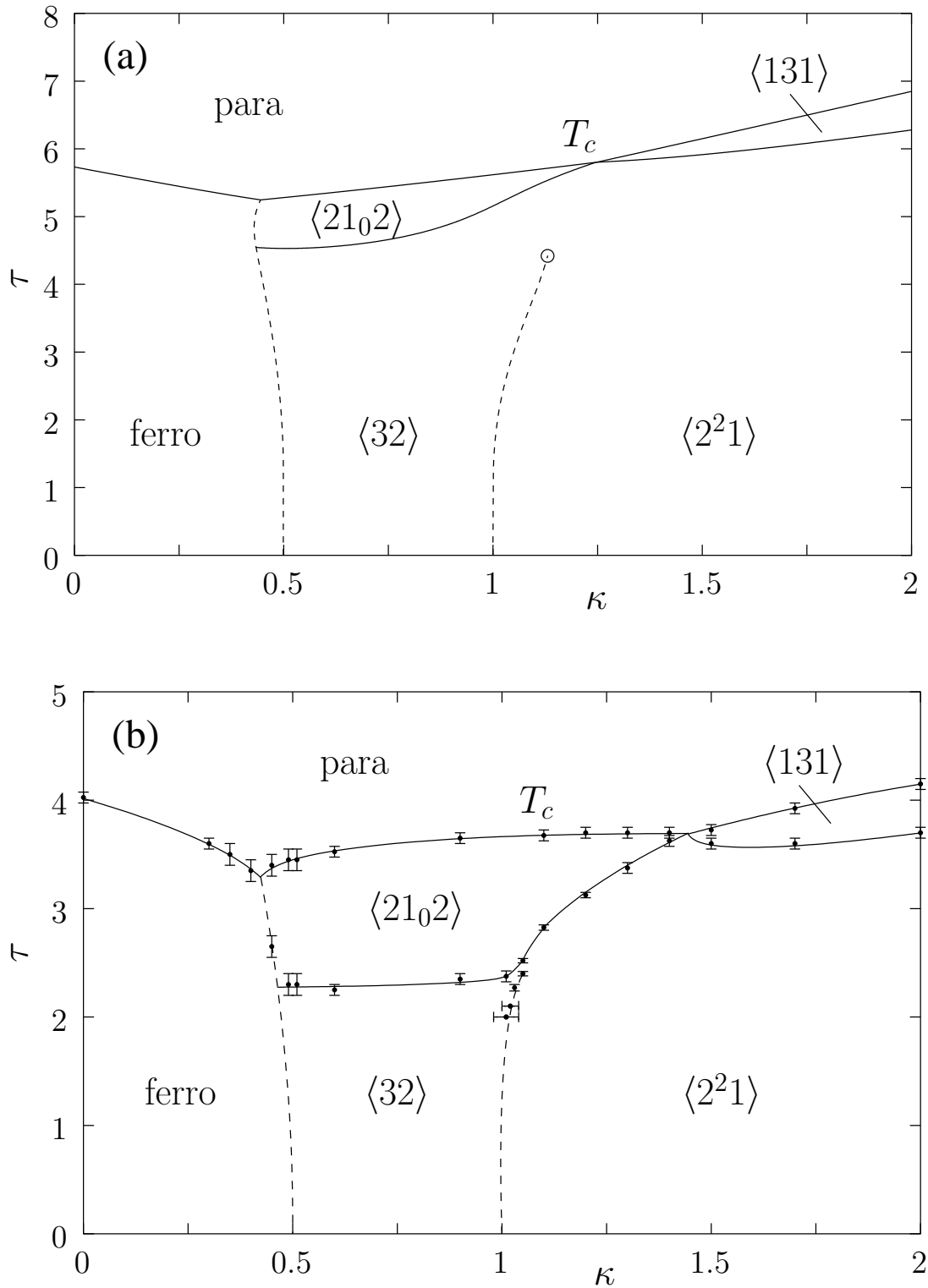

\input{phases_5mf.pstex_t}\\[1cm]
\input{phases_5mc.pstex_t}
\caption{Phase diagram of the ANNNI model for $L= 5$ in
the $(\tau= k_BT/J_0, \kappa= -J_2/J_1)$--plane using
(a) mean--field theory and
(b) Monte Carlo simulations. Dashed (solid) lines denote
transition lines of first (second) order. In the Monte
Carlo case, the lines are guides to the eye.}
\end{figure}

\begin{figure}
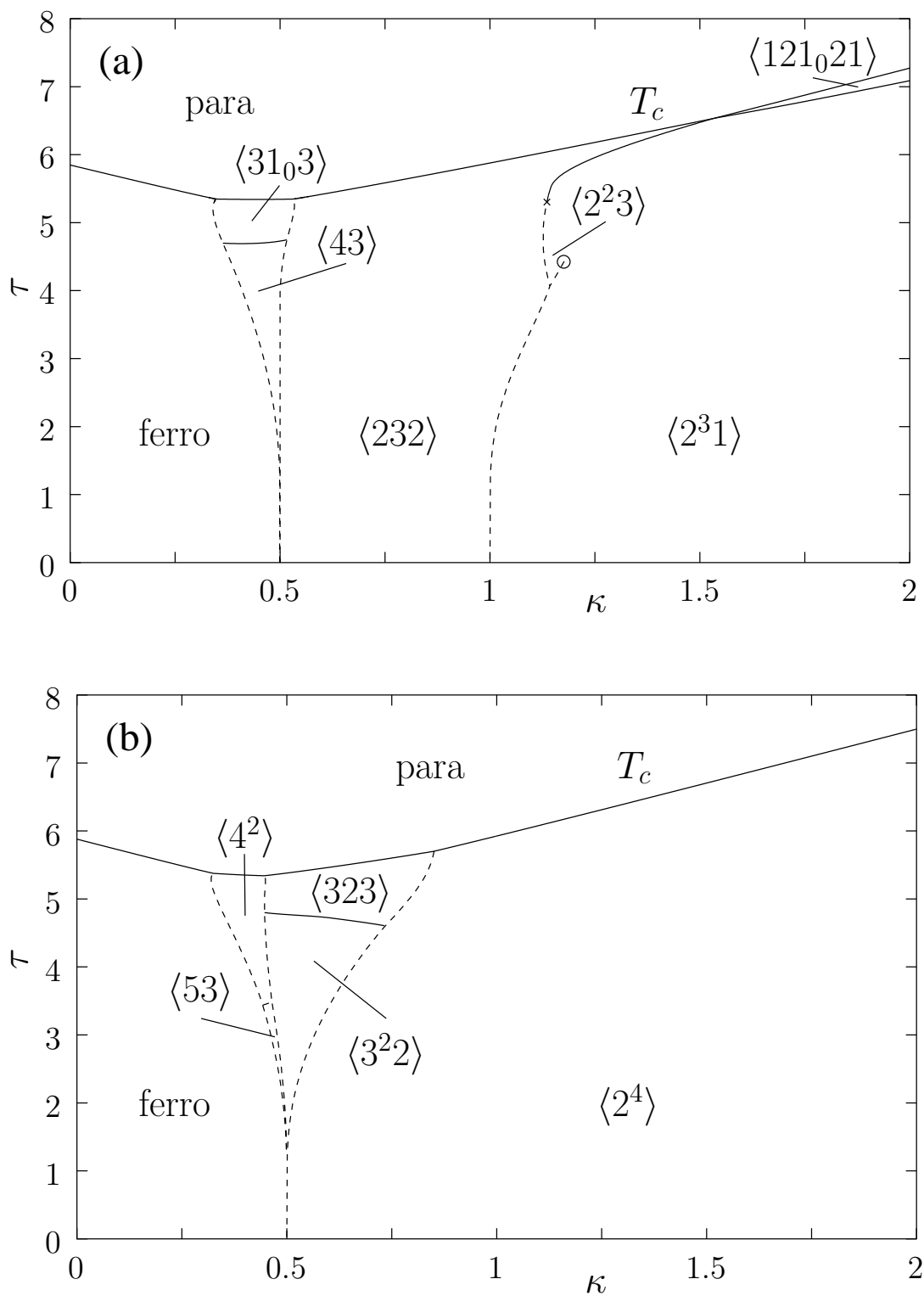

\input{phases_7mf.pstex_t}\\[1cm]
\input{phases_8mf.pstex_t}
\caption{Phase diagrams of the ANNNI model, using mean--field theory,
for thin films with (a) $L=7$ and (b) $L= 8$ layers.}
\end{figure}

Examples of entire phase diagrams are depicted in Figs. 1 and 2 (a more
complete set of phase diagrams can be found elsewhere \cite{Catrein}). In
general, for thin films, there are not many ordered phases. The
structures close to the transition to the paramagnetic
phase, $T_c$, may be obtained in mean--field approximation without
difficulty from the 
linearized theory. One finds that the pattern of
the layer magnetization $m(z)$ is either
symmetric or antisymmetric about the center of the film. 
Accordingly, some of the low temperature phases discussed above
cannot extend up to $T_c$, e. g. the $\left< 43 \right>$ phase (note
that mean--field theory seems to reproduce the low temperature
behaviour qualitatively correctly). Each of the symmetric
or antisymmetric phases is usually stable in a rather
wide region of the phase diagram, for
the thin films we considered. These features, symmetrization
and possible instability of low temperature phases close to $T_c$, are
in marked contrast to the mean--field description of
the ANNNI model with $L \longrightarrow \infty$, where
the structures below $T_c$ form a devil's staircase. Perhaps
most interestingly, we identified a new type of antisymmetric
phase, with the center layer of the film
being disordered, for odd $L$. Examples
are the $\langle 11_01 \rangle$, $\langle 21_02 \rangle$, 
$\langle 31_03 \rangle$, and $\langle 41_04 \rangle$ phases for
$L$= 3, 5, 7, and 9 (we modified the standard low temperature
notation in an obvious way by denoting the disordered layer
by "$1_0$", i.\ e.\ for instance, $\langle 11_01 \rangle$ stands
for a magnetization pattern
where $m(1)> 0$, $m(2)=0$, and $m(3)< 0$). For $L= 9$, at
larger ratio of the competing interactions, $\kappa$, one finds
another partially disordered phase, namely
the $\langle 221_022 \rangle$ phase.

Partially disordered phases, with period arrangements of disordered
layers, occur in mean--field
theory of the ANNNI model, $L \longrightarrow \infty$, at
sufficiently weak intra--plane
couplings, $J_0$ \cite{Salinas,Nakani}. They turned out to be an
artefact of the approximation, as seen in Monte Carlo
simulations \cite{Rotthaus}. However, in
the present case, simulations confirm the prediction of
mean--field theory of disordered center layers, as we checked
for $L$ up to 9. Obviously, those phases are expected to exist only
above the critical temperature of the two--dimensional Ising
model, see Fig. 1.

Note that the sign of the magnetization $m(z)$ may change, at
higher temperatures, in some layers without any phase transition. To
characterise such a phase, we based the notation, used
in the phase diagrams, Figs. 1 and 2, on
the magnetization pattern of that phase in the low temperature region. An
example is the $\langle 323 \rangle$ phase, see Fig. 2, where, close
to $T_c$, the magnetization in the third and sixth layers may change
sign so that the 2--band in the center expands, formally, to a 4--band, with
the 3--bands shrinking to 2--bands. However, the phase keeps
its symmetry, and there is no transition or degeneracy associated
with this change of sign.
 
At intermediate temperatures, interesting features are observed as well. 
In mean--field theory, the
transition line to the $\langle 2^{(L-1)/2}1 \rangle$ phase at
$\kappa \approx 1$, being of first order at low temperatures, ends
in a critical point ($L= 5, 7$, and 9). Simulations, for $L=5$, seem
to be in qualitative agreement, see Fig. 1, but careful and extensive
finite--size analyses are needed to establish this aspect definitely.
Moreover, at intermediate temperatures additional phases may
be stabilized, springing neither directly from the ground states
nor being (anti--)symmetric. An example is the $\langle 54 \rangle$ phase for
$L= 9$ (undergoing, at higher temperatures, a transition
to the symmetric $\langle 41_04 \rangle$ phase). Indeed, if $L$ gets large
enough, structure combination branching processes may happen, when
being compatible with the
film thickness (again, with, possibly, slight rearrangements
and modifications of bands, especially near the surface, as in the
low temperature region).
   
In summary, most of the distinct commensurate
phases of the  ANNNI model in the limit
$L \longrightarrow \infty$
are suppressed in thin films. Perhaps
surprisingly, however, modified structures and novel
phases are induced by
the (small) thickness of the film, $L$, for example, among the phases
springing directly from the multiphase point and among those
close to the paramagnetic transition line. Certainly, it
may be interesting to study the robustness of
the findings against varying model parameters such as, for
instance, the intralayer coupling, $J_0$, in the surface planes.

\ack
Useful discussions with J. Hager are
gratefully acknowledged.

\end{document}